# Enhanced knowledge retention through MedScrab: an interactive mobile game

Don Roosan[a], Tiffany Khao[a,&], Huong Phan[a,&], Yan Li[b,&]
[a]Department of Pharmacy Practice and Administration, Western University of Health Sciences, 309 E 2nd st, Pomona, California 91766, USA
[b]Center for Information Systems and Technology, Claremont Graduate University, 150 E 19th st, Claremont, California 91711, USA

**Abstract.**
Noncompliance with medication regimens poses an immense challenge in the management of chronic diseases, often resulting in exacerbated health complications and recurrent hospital admissions. Addressing this gap, our team designed an innovative mobile game aimed at bolstering medication adherence and information retention within the general population. Employing Amazon Mechanical Turk, participants were enlisted and allocated into two cohorts: one engaged with our mobile game and the other perused an informational pamphlet about medication. Both cohorts underwent a pre-intervention quiz, followed by their respective interventions, and concluded with a post-intervention quiz. Primary outcome measures included the difference in quiz scores and the game play duration. The investigation encompassed 243 participants with homogenous baseline attributes. Participants interacting with the mobile game depicted a significant enhancement in their post-intervention scores compared to the pre-intervention scores (1.83 ±1.187 vs 2.92 ±1.696, $p < 0.001$). We observed a notable correlation of 0.346 ($p<0.001$) with a robust medium effect size of 0.641 (0.503 - 0.779). Although the duration of game play and post-intervention scores didn't exhibit a direct correlation, a tendency towards superior post-intervention scores was evident among participants who dedicated more time to the game. The interactive mobile game we developed exhibits potential as an engaging instrument for empowering patients and caregivers. Providing critical medication information and the potential side effects in a manner that increases retention would thereby mitigate medication noncompliance. Future research endeavors should focus on optimizing and broadening the application of such mobile interfaces to fortify public health initiatives.

1. **Introduction:**

Medication nonadherence affects up to 50% of patients with chronic conditions on multiple medications and causes an estimated 125,000 avoidable deaths and $100 billion in hospitalization expenses annually. U.S. Centers for Medicare & Medicaid Services (CMS) uses proportion of days covered (PDC) to measure adherence for patients with chronic conditions.[1–3] Pharmacists provide a Patient Package Insert (PPI), which is a summary of important information about the medications. The FDA requires PPIs to have the following information: name of the drug, name of manufacturer, benefits and risks of the medications, proper use, contraindications, side effects, and instructions on how to manage the side effects.[4] Lastly, even if patients read the PPI and medication leaflets, they may not retain much information. Current research indicates that mobile applications can potentially be utilized as educational tools to help patients learn about their medication.[5,6] Furthermore, mobile games have been shown to effectively promote medication adherence and lead to positive health outcomes amongst different disease states.[7–9] Therefore, new interactive and patient-engaging mHealth applications are needed to educate the general patient population on various medications.

The mobile application we developed is a free, interactive, word-based game designed to provide patients with medication information. All the medication information was collected and organized from tertiary resources such as Lexicomp, manufacturer PPIs, and the USFDA website. In addition, the data has been verified by multiple professional pharmacists. Our mobile game has five different levels for each medication. Each level starts with lookalike Scrabble word games. There is a multiple-choice question to assess the user's knowledge about the drug before proceeding to the next level. Once all levels are completed, there is a final quiz that consists of multiple-choice questions to check the general knowledge of medication. Overall, the game has a novel approach to the delivery of complex drug information to the general patient population. Additionally, the game application is free to download in the app store, making it easily accessible to patients and their caregivers.

Due to the unique designs and features, we hypothesized that playing our mobile game can improve medication information recall and medication adherence. To test our hypothesis, we proposed an experimental design study to evaluate the effectiveness of the mobile game in improving medication information retention.

2. **Materials and Methods:**

*2.1 Study settings:*
This study received the approval of the Institutional Review Board of a large university. All study subjects were informed of the procedure and provided written consent to participate. Assuming a statistical power of 80% and a 95% confidence interval with a 2-sided test, an estimated 99 participants were needed for the study.

*2.2 Study design:*
Our game is a web-based and smartphone-based videogame application. The application was designed similarly to a Scrabble word game that helps users learn general information about the medications. In this study, participants were given 40 minutes to download, sign-up, and play the mobile game associated to the assigned medication. The duration of playing time was recorded and the participants did not have to use the entire allotted time. In the end, participants completed a post-game quiz related to the assigned medication.

*2.3 Study participants and recruitment:*
The participants were recruited from Amazon Mechanical Turk (MTurk), a crowdsourcing marketplace that asks participants to perform tasks over the Internet. These participants were screened online with our preset criteria before entering the rest of the online experiments. The requirements to participate in the study included:
1. Participant must be 18 years old or older
2. Participant can read and understand English.
3. Participant must be a Mturk worker living in the United States.

The worker's HIT (human intelligent task) approval rating must be greater than 95%, and the worker's number of HITs approved must be greater than 1000. Anyone who is less than 18 years old, does not have access to a mobile device (smartphone or tablet), refuses to sign the written consent, or has an English comprehension level of less than 8th grade was excluded. Mturk workers who did not have a US IP address, or did not have approval ratings of greater than 95%, and at least 1000 HITs were also excluded.

*2.4 Demographic variables:*
Demographic variables include participants' age, gender, educational background, race, and healthcare-related career or education. All information was gathered using a survey-style questionnaire, as stated in the consent form regarding the data we collected in the study.

*Primary Outcome variables:*
Outcome variables include pre-quiz scores, post-quiz scores including one attention question, and duration of time spent playing the game. Mturk workers were required to complete either a game task on our application or a reading task, which was embedded in a survey. To ensure that workers were not "gaming" our project, a minimum of 3 minutes and 6 minutes spent completing the game and reading tasks, respectively, was a requisite to receive compensation. This data was collected via quizzes with multiple-choice answer questions. These questions were related to the participants' assigned medication group, either simvastatin or sertraline. Our study focused on simvastatin and sertraline as they are commonly prescribed and can be used to treat and prevent a multitude of diseases.[10,11]

*Procedure:*
Participants were screened and recruited online from Amazon Mechanical Turk (MTurk). The participants in the intervention group were presented information regarding a particular drug (i.e., simvastatin or sertraline) through playing the mobile game (game task). The participants in the control group acquired the drug information through reading a pamphlet (reading task). These tasks were embedded in two of our four surveys. Part 1 of the experimental design study included four surveys. Participants were paid $1.00 to complete any one of the four surveys on the Mturk platform. Each participant was allowed to complete only one survey and thus be excluded from doing the other three surveys. Part 2 of the study occurred seven days later and included two follow-up surveys for eligible participants who previously completed part 1. Firstly, included participants were randomized by MTurk to one of the four surveys. Embedded in each of the four experimental design surveys was a link to either: a simvastatin medication guide, a sertraline medication guide, the simvastatin game, or the sertraline game. Each survey contained eight questions specific to the given drug, either simvastatin or sertraline (3 pre-knowledge questions and 5 multiple choice post-quiz questions), 1 adherence question with an optional follow-up free response question regarding adherence, 1 preconceived notion question, and 1 attention check question embedded in the multiple-choice post quiz to detect gaming behaviors. Lastly, the survey concluded with 6 demographic questions. Workers were required to input their Mturk worker ID to generate a random Qualtric code once the survey was completed. Workers were instructed in the survey to cut and paste the code into the Mturk website. After taking the pre-quiz, participants were given 40 minutes to download, sign-up, and play the game for the assigned medication group, 10 minutes to complete a post-quiz and 5 minutes to complete the demographic survey. To ensure high quality data, the administrators reviewed the app's backend activity log, email address, completion status, and duration spent playing the game. We set a time restriction

that only allows a worker to submit when the minimum duration of time has elapsed. Participants who completed the first part were categorized into two groups: workers who completed the simvastatin quiz and workers who completed the sertraline quiz. These participants were invited to complete the second part and were compensated $2. The questions of the second part were identical to the post-quiz of part 1.

### 3. Results

Two hundred forty-three individuals have participated in this study. Since this is a controlled before and after study, participants' demographic and baseline clinical characteristics were not significantly different. These participants were 57% males and 43% females. Also, 79% of the participants were white. Their ages were between 18 and 75 years (44% were 26-35 years old, 26% were 37-45 years old), of which 53% were undergraduate education level, and 27% were master's education level.

**Outcomes:**

Among those 243 subjects, there was a significant difference between pre-scores and post-scores after playing the mobile game (1.83 ±1.187 vs 2.92 ±1.696, p <0.001) with a positive correlation of 0.346 (p<0.001) and a medium effect size of 0.641 (0.503 – 0.779) which is demonstrated in table 1.

**Table 1** Our study results show that participants got higher score after playing the mobile game compared to before they play the game. Also, the more time participants spent the time playing game, the higher score they received for post quiz.

| Outcomes | Pre-score (±SD) | Post-score (±SD) | Mean different (95%CI) | P-value | Correlation | Effective size |
|---|---|---|---|---|---|---|
| Quiz Scores | 1.83 ±1.187 | 2.92 ±1.696 | -1.091 (-1.305-0.876) | p <0.001 | 0.346 (p<0.001) | 0.641 (0.503-0.779) |

| Outcomes | | High playing time (±SD) | Low playing time (±SD) | Mean different (95% CI) | P-value |
|---|---|---|---|---|---|
| Post Quiz Score | Simvastatin | 3.86 ±1.489 | 2.85 ±1.941 | 1.006 (0.351-1.661) | P=0.001 |
| | Sertraline | 2.47 ±1.475 | 1.98 ±1.390 | 0.491 (-0.036-1.017) | p= 0.034 |

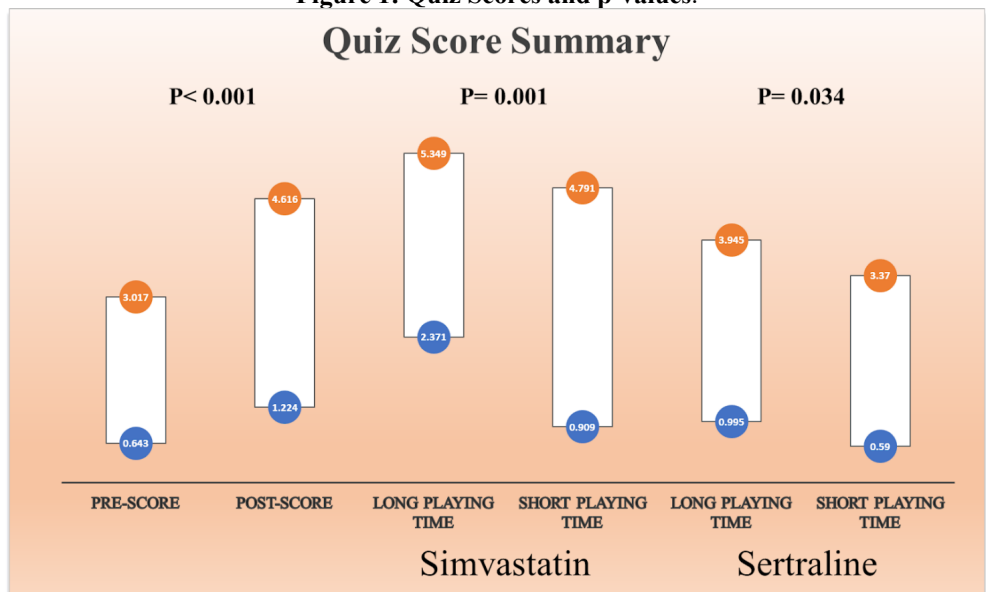

**Figure 1: Quiz Scores and p-values**.

There was no correlation between playing duration and post-score (Kendall's Rank Correlation 0.07, p= 0.133 and Spearman's Rank Correlation 0.094, p= 0.142). However, among 119 subjects who received questions concerning simvastatin, people who spent more time playing the game obtained a better post-score than those who spent less time playing (3.86 ±1.489 vs. 2.85 ±1.941, p= 0.001). Similarly, among 124 subjects who received questions concerning sertraline, people who spent more time playing the game also obtained a better post-score than those who spent less time playing (2.47 ±1.475 vs. 1.98 ±1.390, p= 0.034).

## 4. Discussion:

mHealth application research has grown exponentially in recent years due to widespread smartphone usage.[12] Our gaming application was designed for the general population and is capable of acquiring usage data, such as duration of play, during experimental research, which may provide a new perspective for future mHealth game innovations. Our study reveals that people who spent more time playing the mobile game scored better on the post-quiz than those who spent less time and suggest that it can better help people recall medication information.

Gaming applications can also assist dementia patients and their caregivers in achieving better medication management. Alzheimer's disease and related dementia (ADRD) is a chronic condition that deteriorates people's cognitive functions and impairs their ability to complete daily activities. According to the WHO Global Status Report 2021, the number of people living with dementia could increase from 55 million in 2020 to 139 million in 2050.[13] The majority of these patients receive care from caregivers who suffer from increased stress and severe depressive symptoms, amongst other health problems, in comparison to caregivers for non-dementia patients.[14,15] This is often referred to as caregiver burden, which can be defined as a high level of physical, psychological, emotional, behavioral, and financial burden that may be experienced by dementia caregivers.[16] To address this problem, many mobile applications have been developed as accessible tools for caregivers to learn about their patient's condition and medication information.[17] Similar to other mHealth applications, our mobile game can help educate caregivers obtain essential information regarding the patient's medications such as indications, side effects, and management.

Moreover, the mobile game we developed can serve as a training tool for ADRD patients.[18] Our mobile application has the potential to stimulate multiple parts of the brain for ADRD patients, helping to prevent the deterioration of their memory. In addition to improving memory and cognition, playing mobile games can enhance eye-to-hand coordination and better improve the attention of users. Therefore, our mobile application can be an educational tool for both patients and caregivers to learn about the medications, better manage the conditions, and improve the overall quality of care.

Our study results indicate that participants who played the mobile game longer received higher scores on the post-game quiz and retained more medication information. Furthermore, the mobile application has shown that interactive game-based education can effectively package lengthy, cumbersome drug information into easy-to-digest information for the general patient population. In addition, our mobile game is a free application which patients can easily carry to play and learn about drug information at their convenience.

Nevertheless, our mobile game does not aim to replace traditional pamphlets and teaching methods, but instead gives patients another option to learn about overwhelming drug information. Our mobile game can provide some insight to spark creativity in future digital health development. Future studies may need to consider perspectives from a broader population to better achieve user-centered design and thus improve health outcomes. We hope individuals and organizations, including general users, healthcare providers, and government entities can promote our free application along with other creative applications to help the general population gain drug information interactively.

## 5. Conclusion:

We successfully developed a web-based and smartphone-based videogame mobile application, which helps users to learn drug information in an engaging way. In this study, people were divided into two different groups: either simvastatin or sertraline. In both groups, people who spent more time playing the mobile game recalled drug information significantly better than those who spent less time. Our mobile game is a free application that can help the general patient population learn about drug information in a fun and effective way.